\documentclass[prl,twocolumn,aps,superscriptaddress,preprintnumbers,letterpaper]{revtex4}
\usepackage{amsmath,amssymb}
\usepackage{epsfig}
\usepackage{graphicx}
\usepackage{amsmath}
\usepackage{amsfonts}
\usepackage{epstopdf}
\newcommand{\be}{\begin{equation}}
\newcommand{\ee}{\end{equation}}
\newcommand{\bear}{\begin{eqnarray}}
\newcommand{\eear}{\end{eqnarray}}
\newcommand{\ba}{\begin{array}}
\newcommand{\ea}{\end{array}}

\begin{document}

\title{Chiral Disorder and Random Matrix Theory with Magnetism}

\author{Maciej A. Nowak}
\email{nowak@th.if.uj.edu.pl} \affiliation{M. Smoluchowski Institute
of Physics and Mark Kac Center for Complex Systems Research,
Jagiellonian University, PL--30--059 Cracow, Poland}

\author{Mariusz Sadzikowski}
\email{sadzikowski@th.if.uj.edu.pl} \affiliation{M.
Smoluchowski Institute of Physics,  Jagiellonian University,
PL--30--059 Cracow, Poland}

\author{Ismail Zahed} \email{zahed@tonic.physics.sunysb.edu}
\affiliation{Department of Physics and Astronomy, Stony Brook University, Stony brook, New York 11794-3800, USA}
\date{\today}

\date{\today}

\begin{abstract}
We revisit the concept of chiral disorder in QCD in the presence of a QED magnetic field $|eH|$.
Weak magnetism corresponds to $|eH|\le 1/\rho^2$ with $\rho\approx 1/3\,{\rm fm}$ the vacuum instanton size,
while strong magnetism the reverse. Asymptotics  (ultra-strong magnetism) is in the realm of perturbative QCD.
We analyze weak magnetism
using the concept of the quark return probability in the diffusive regime of chiral disorder. The result is in
agreement with expectations from  chiral perturbation theory. We analyze  strong and ultra-strong magnetism
in the ergodic regime using random matrix theory including the effects of finite temperature. The strong magnetism
results are in agreement with the currently reported lattice data in the presence of a small shift of the Polyakov
line. The ultra-strong magnetism results are consistent with expectations from perturbative QCD. We suggest a chiral
random matrix effective action with matter and  magnetism to analyze the QCD phase diagram near the critical
points under the influence of magnetism.
\end{abstract}

\maketitle

\setcounter{footnote}{0}

\vskip0.2cm

\section{Introduction}

Chiral quarks in the QCD vacuum break spontaneously chiral symmetry. Key contributors
to this spontaneous breaking are instanton and antiinstanton fluctuations with left handed and
right handed zero modes attached to them~\cite{DIAKONOV,Schafer:1996wv,book} (and references therein).
The random nature of the instanton and antiinstanton
fluctuations cause these zero modes to spread near zero virtuality. This fundamental property of the
QCD quark spectrum is captured by the Banks-Casher relation~\cite{Banks:1979yr}

\be
\left< {\bf \Psi}_4^\dagger {\bf \Psi}_4\right> \equiv  \Sigma_4=\pi\rho_4(0)
\label{001X}
\ee
that ties the quark condensate to the quark spectral density $\rho_4(\lambda)$ at zero virtuality.
Note the positive convention for the chiral condensate.
In the QCD vacuum with instanton density ${\bf n}_4$ and size $\rho$, the quark condensate is expected
to scale as $\Sigma_4\approx {\bf n}_4\rho$.
Using the Gell-Mann Oakes Renner relation $m_\pi^2F_\pi^2=2m\Sigma_4$, (\ref{001X}) is just
the analogue of the Kubo formula for the DC conductivity in metals

\be
\sigma_C=F_\pi^2/\pi=D\rho_4(0)
\label{002X}
\ee
The pion decay constant $F_\pi$ defines the chiral conductivity, with the chiral diffusion
constant $D=F_\pi^2/\Sigma_4$.
Through (\ref{001X}-\ref{002X}) chiral quarks trapped in  an Euclidean 4-volume $V_4=L^4$
behave much like electrons in disordered metallic grains with $\lambda_T=D/L^2$
playing the role of the Thouless energy~\cite{Janik:1998ki,OSVERBAAR,TAKEH,BERBE}. Quark with virtualities
$\lambda<\lambda_T$ are in the ergodic regime and are well described by random
matrix theory, while quarks with virtualities  $\lambda>\lambda_T$ are in the diffusive regime
which is amenable to chiral perturbation theory.
In the presence of a fixed external magnetic field $H$ the chiral disorder is affected.
In this paper we address the details of these changes.

Weak magnetism is mainly affecting
the diffusive properties of chiral fermions through a renormalization of the
low energy parameters, e.g. $\Sigma_4, F_\pi,  m_\pi$. Although the chiral expansion in
the massless limit relies on $|eH|/(4\pi F_\pi)^2<1$ and therefore suggests that weak
magnetism operates in the realm of $eH|\leq 1\,{\rm GeV}^2$, the validity range is substantially
smaller. Indeed, since instantons may be  at the origin of the spontaneous breaking
of chiral symmetry we expect weak magnetism to break down when the instanton
size $\rho$ is resolved. As a result, the range of weak magnetism is
$|eH|\leq 1/\rho^2\approx 1/3\,{\rm GeV}^2$ and will be referred to as diffusive.

Strong magnetism corresponds to $|eH|> 1/\rho^2\approx 1/3\,{\rm GeV}^2$ with a magnetic field starting to dwarf the
chromomagnetic field carried by the instanton. This range will be referred to as ergodic.
As a result, light and chiral quarks in the QCD
vacuum switch from locking their spin to the instanton color to lining their spin with the external magnetic field.
They form Landau orbits which are gapped by $|eH|$ near zero virtuality.
The exception is the lowest Landau level (or LLL for short),  where the magnetic contribution cancels
the zero point motion. As a result, the LLL with its huge degeneracy piles up at zero virtuality becoming
a potential contributor to the spontaneous breaking of chiral symmetry. The LLL is inherently 2-dimensional.
This phenomenon of dimensional reduction from 4 to 2 in Euclidean space with accumulation near zero virtuality
of the LLL favors the spontaneous breaking of chiral symmetry through any residual interaction.
This phenomenon is known as magnetic catalysis~\cite{Shovkovy:2012zn} (and references therein).
For instance, residual longitudinal interactions in the LLL through
the  instantons  will cause the spontaneous breaking of chiral symmetry,

\be
\Sigma_4=|eH|\Sigma_2\approx |eH|\sqrt{n_4}\rho
\label{003X}
\ee
with typically ${\bf n}_4\approx 1/{\rm fm}^4$ the vacuum instanton density.

Ultra-strong magnetism is solely characterized by Landau levels with huge degeneracy and small sizes.
As a result, many Landau orbits can fit in a single instanton making the concept of instanton zero modes
not particularly useful. In many ways, this picture resembles that of the initial color glass condensate
with many wee partons saturating the transverse size of a colliding nucleus~\cite{McLerran:2003yx}
(and references therein).
Perturbation theory becomes the rule and will cause the LLL to spread near zero virtuality.
The spontaneous breaking of chiral symmetry through perturbative gluons follows with

\be
\Sigma_4=|eH|\Sigma_2\approx |eH|^{3/2}
\label{004X}
\ee
as expected from dimensional arguments. We expect this scaling to take place
for $|eH|\geq 10/\rho^2\approx 3\,{\rm GeV}^2$ or when about 10 Landau orbits can fit
within the instanton transverse size thereby dwarfing the zero modes.

In section 2, we discuss weak magnetism in the diffusive regime of 4-dimensional chiral quarks
trapped in $V_4$ using the concept of the quark return probability. The result is in line with the
one from  chiral perturbation theory. In sections 3 and 4, we discuss ultra-strong and strong magnetism
in the ergodic regime where the instanton size is resolved through the use of
random matrix theory. The results are in line with the expectations of dimensional reduction
from the LLL. In section 5 we compare our results to current lattice simulations~\cite{Bali:2013cf}.
In particular we show that a small shift in the trivial Polyakov holonomy is magnified by the
chiral transition and may account for the anti-catalysis observed on the lattice as recently
suggested in~\cite{Bruckmann:2013oba}.
In section 6, we show how random matrix models for QCD with magnetism can be
constructed to shed light on the comparison with widely used constituent quark models of the NJL type,
and help analyze a number of issues in the QCD phase diagram with magnetism.
Our conclusions and prospects are in section 7.

\section{Diffusive Regime with Weak Magnetism}

The dynamics of QCD light quarks in a 4-dimensional Euclidean box is chiral and diffusive in the long
wavelength limit.  The diffusive nature is best captured by the quark return probability~\cite{Janik:1998ki}

\be
P(0,\tau)=\left<\left|u^+(\tau) u(0)+d^+(\tau) d(0)\right|^2\right>
\label{00}
\ee
for 2 light flavors. In the absence of magnetism, the vacuum is isospin symmetric and (\ref{00})
is dominated by the the triplet of pions in an Euclidean box

\be
P(0,\tau)=2\left(P_0(0,\tau)+P_\pm (0,\tau)\right)
\label{000}
\ee
with

\be
P_{0,\pm}(0,\tau)\approx \sum_Q e^{-DQ^2 |\tau|}
\label{001}
\ee
for a triplet of massless pions.   The sum is over the isotriplet
of charged pions  or diffuson modes with momenta $Q_{\mu}=n_{\mu}2\pi/L$
in a periodic $V_4=L^4$ Euclidean box. The quark return probability (\ref{00})
in the chiral limit obeys a sum rule

\be
\Sigma_4=-\lim_{m\to0}\lim_{V_4\to\infty}\frac{1}{V_4} \int_0^{\infty} P(0, \tau)\,d\tau
\label{002}
\ee

In the presence of magnetism, the vacuum is no longer isospin symmetric. As a result
the free chargeless pion $\pi^0$ remains gapless, while the free charged pions $\pi^\pm$
are gapped in Landau levels (LL). For a constant magnetic field with $A^1=-Hx_2$, the
LL are

\be
\lambda_\pm^2=|eH|(2n+1)+p_3^2 +p_4^2 +m^2_\pm
\label{003}
\ee
with degeneracy $\phi=eHL^2/2\pi$.  The change in the quark return probability follows from
the change in the diffuson propagator for the charged pion modes

\be
P_\pm(H, \tau)=\sum_{n,m,k} \frac{eHL^2}{2\pi} e^{-D \tau[(n+1/2) 2eH +(m^2+k^2)(2\pi/L)^2 +m_\pm^2] }
\label{004}
\ee
The change in the quark return probability is the change in the charged diffuson modes and is
captured by the difference

\be
I=\int_0^{\infty}[P_\pm (H,\tau)-P_\pm (0,\tau)]d\tau
\label{005}
\ee
In the chiral limit, replacing the sums over free momenta by integrals and summing explicitly over the
Landau levels of the charged pions (i.e. the  diffusons) we obtain

\be
I=\frac{eHV}{16\pi^2D}\int_0^{\infty} \left(     \frac{1}{z \sinh z} -\frac{1}{z^2} \right)dz = -\frac{eHV}{16\pi^2 D} \ln 2
\label{006}
\ee
Using the  value of the diffusion constant we arrive at
\be
\Sigma_4(H)=\Sigma_4(0) \left(1+\frac{e H \ln 2}{16 \pi^2 F_{\pi}^2}\right)
\label{007}
\ee
which is the result of chiral perturbation theory in leading order~\cite{Shushpanov:1997sf}. As expected, the diffusive
regime is the regime of chiral perturbation theory albeit in a finite Euclidean box~\cite{Janik:1998cg}. Chiral
perturbation theory in a finite box was formulated systematically in~\cite{Gasser:1987ah}.

\section{Ergodic Regime with Ultra-Strong Magnetism}

For strong magnetism the chiral disorder in 4 dimensions with an isotriplet of pions transmute to a chiral
disorder in quasi-2-dimensions with a chargless pion, as the charged diffusons become gapped.
The transmutation takes place when the magnetic field resolves the instanton size or
$|eH|>1/\rho^2\approx 1/3\,{\rm GeV}^2$.  In this regime, magnetism dwarfs the instanton chromo-magnetism
with Landau levels becoming the lore. The LLL dominates the infrared physics near zero virtuality
triggering the catalysis of chiral symmetry breaking by magnetism.

Indeed, for strong magnetism the free quark spectrum in 4-dimensions is given by Landau levels

\be
\lambda^{\pm}_{n,s}(\omega,k_z) =\pm \sqrt{\omega^2+k_z^2+|eH|(2n+1-s)+m^2}
\label{9}
\ee
with current quark masses. For simplicity we will consider $N_F=1$ with charge $e$ unless specified otherwise.
Each $(\omega, k_z)$ branch is

\be
\phi=|eH|L_xL_y/2\pi\equiv |H|L_xL_y/(h c/e)
\label{X10}
\ee
 degenerate. (\ref{X10}) measures the magnetic flux in units of the quantum flux $hc/e$ and is
 integer valued  on a 4-dimensional lattice with periodic (toroidal) boundary conditions.  The
 Lowest  Landau branches  $n=0,s=1$

\be
\lambda^{\pm}_{0,1}(\omega,k_z) =\pm \sqrt{\omega^2+k_z^2+m^2}
\label{10}
\ee
are paired by chirality and dominate in the infrared limit. They are $2\phi$ degenerate.
The higher Landau levels are gapped at zero virtuality.
We note that at finite temperature the $\omega$-spectrum in (\ref{10}) is dominated by the lowest Matsubara mode
$\omega\rightarrow \omega_0=\pi T$~\cite{Jackson:1995nf}. This approximation will be used to probe critical points in phase diagrams
with magnetism.

While the color interactions are involved within and between Landau levels, a model
independent understanding can be achieved through the Banks-Casher formulae
used earlier. In the presence of strong magnetism the localization of the quark spectrum
in the transverse hyperplane to the magnetic field suggests

\be
\left<\Psi^\dagger_4\Psi_4\right>=\pi\rho_4(0)=\frac{\pi}{V_4}\frac {2\phi}{\Delta\lambda}=\frac {|eH|}{\beta L_z\Delta\lambda}
\label{10X0}
\ee
where $\Delta\lambda$ is the typical level spacing near zero virtuality. In a spontaneously
broken QCD vacuum with no magnetism $\Delta\lambda\approx 1/V_4$.
The last relation in (\ref{10X0}) reflects on the dimensional reduction of the quark spectral
function near zero virtuality from 4 to 2 dimensions,

\be
\rho_4(0)=|eH|\rho_2(0)
\label{10XX}
\ee
and therefore the condensate relation

\be
\left<\Psi^\dagger_4\Psi_4\right>=|eH|\left<\Psi^\dagger_2\Psi_2\right>
\label{11X}
\ee
In the infrared regime, this suggests the field redefinition

\be
\Psi_4(x_0,x,y,z))\rightarrow {\sqrt{|eH|}}\,\Psi_2(x_0,z)
\label{11}
\ee
with the pertinent canonical dimensions made explicit.

The breaking of chiral symmetry in terms of ${\bf \Psi}_2$
is quasi-2-dimensional. It does not upset the Mermin-Wagner theorem. Indeed, the random gluonic interactions
that cause the spontaneous breaking of chiral symmetry and therefore the redistribution
of the 2-dimensional quark spectrum near zero virtuality are still blind to magnetism and fully 4-dimensional.

In quasi-2-dimensions the free quark fields $\Psi_2$ are dominated by the
branches (\ref{10}) and any residual gluonic interaction will cause them to redistribute.
For ultra-strong magnetism with $$|eH|\geq 10/\rho^2\approx 3\,{\rm GeV}^2$$ the interactions among the quarks in the LLL are mainly
perturbative as the instanton quasi-zero modes become obsolete. Assuming the residual
interactions in the infrared for the lowest landau level (LLL) and lowest Matsubara frequency
to be totally random but chiral, the partition function for these modes is best captured by a new chiral RMM

\begin{eqnarray}
&&{\bf Z}_{LLL}(m,T,\phi)=\nonumber\\&& \int\,d{\bf A}\,e^{-(\phi/|eH|){\rm Tr}\left({\bf A^\dagger A}\right)}
\,\,{\rm det}
\left|\begin{matrix}
m & i\omega_0+i{\bf A}\\
i\omega_0+i{\bf A}^\dagger& m\\
\end{matrix}\right|\nonumber\\
\label{12}
\end{eqnarray}
where ${\bf A}$ is a $\phi\times \phi$ real matrix with a Gaussian distribution. Recall that we have set $N_F=1$.
The novelty of (\ref{12}) is in the scaling of the Gaussian distribution of the matrix elements
with magnetism. It captures the
random interactions induced by plain perturbative
gluons between the $\phi$ degenerate quarks ${\bf \Psi}_2$ in the LLL
and amounts to the 4-fermi interaction
$$|eH|\left({\bf \Psi}_2^\dagger{\bf \Psi}_2\right)^2$$
as expected from dimensional reduction.
The Gaussian weight enforces a level spacing between the eigenvalues of the
chiral matrix to be of order

\be
\Delta\lambda\sim\frac{\sqrt{|eH|}}{\phi}
\label{12X}
\ee
By the Banks-Casher formulae we expect the chiral condensate in a symmetric box $V_4=L^4$

\be
\left<\Psi^\dagger_4\Psi_4\right>=\pi\rho_4(0)=\frac{\pi}{V_4} \,\frac{2\phi}{\Delta\lambda}
\sim\frac 1{2\pi}\, |eH|^{3/2}
\label{12XX}
\ee
in agreement with the dimensional arguments presented in (\ref{004X}). This result is born out by explicit calculations as we now show.

The quark condensate follows through (\ref{11X}) as

\be
\left<\Psi^\dagger_4\Psi_4\right>=\lim_{m\to 0}\frac{|eH|}{\beta L_z}\frac{\partial{\rm ln}{\bf Z}_{LLL}}{\partial m}
\label{13}
\ee
In the large $\phi$ limit, the partition function in (\ref{13}) is analyzed through standard arguments by fermionizing the
determinant in (\ref{12}), performing the Gaussian integral and then bosonizing the resulting 4-fermi interaction
using the Hubbard-Stratonovitch transform,

\begin{eqnarray}
&&Z_{LLL}(m,T,\phi)=\nonumber\\&&\int\,d{\bf b}\,e^{-\frac{\phi}{|eH|}{{\bf b}^\dagger{\bf b}}}\
{\rm det}^\phi
\left|\begin{matrix}
{\bf b}+m & i\omega_0\\
i\omega_0& {\bf b}^\dagger+m\\
\end{matrix}\right|
\label{BZ1}
\end{eqnarray}
Inserting (\ref{BZ1}) in (\ref{13}) yields

\begin{eqnarray}
\left<{\bf \Psi}_4^\dagger{\bf \Psi}_4\right>=\lim_{m\to0}\lim_{L_z\to\infty}
\frac{|eH|}{\beta L_z}\left<{\rm tr}\,\left|\begin{matrix}
{\bf b}+m & i\omega_0\\
i\omega_0& {\bf b}^\dagger+m\\
\end{matrix}\right|^{-1}\right>\nonumber\\
\label{Z3}
\end{eqnarray}
The averaging is carried using the measure in (\ref{BZ1}). In the saddle point approximation, the result is

\be
\left<\Psi^\dagger_4\Psi_4\right>=\frac{|eH|}{\beta L_z}\,\frac{2\phi}{\sqrt{|eH|}}\sqrt{1-\omega_0^2/|eH|}
\label{14}
\ee
which can be re-written as

\be
\left<\Psi^\dagger_4\Psi_4\right>=\frac{L_xL_y}{\beta L_z}\,\frac 1{\pi}\,|eH|^{3/2}\sqrt{1-\omega_0^2/|eH|}
\label{15}
\ee
which is (\ref{12XX}) for a symmetric box. The critical temperature is field dependent

\be
T_c=\frac{\sqrt{|eH|}}{\pi}
\label{16}
\ee

Ultra-strong magnetism implies tightly paired chiral quarks in the vacuum.
The RMM results (\ref{12XX}) and (\ref{16})
are consistent with dimensional arguments for asymptotically strong magnetism,
and the perturbative QCD arguments presented in~\cite{Shushpanov:1997sf}.
This analysis ignores the back-reaction of the quarks on the gauge configurations
in the saddle point approximation. The back-reaction maybe important as we detail below.

Finally, we note that a more general RMM for all  Landau
levels can be written similarly by insisting on the free Landau level spectrum for ${\bf A}=0$. Specifically

\begin{eqnarray}
&&{\bf Z}_{LL}(m,T,\phi)=\int\,d{\bf A}\,e^{-(\phi/|eH|){\rm Tr}\left({\bf A^\dagger A}\right)}\prod_{n=0}^\infty\prod_{s=\pm1}
\nonumber\\
&&\times {\rm det}
\left|\begin{matrix}
m & i\omega_0+\epsilon(n,s)+i{\bf A}\\
i\omega_0-\epsilon(n,s)+i{\bf A}^\dagger& m\\
\end{matrix}\right|\nonumber\\
\label{16X}
\end{eqnarray}
with the free Landau level eigenvalue-spectrum

\be
\omega_0^2+\epsilon^2(n,s)+m^2=\omega_0^2+|eH|(2n+1-s)+m^2
\ee
(\ref{12}) is (\ref{16X}) restricted to the LLL with $n=0, s=1$. Other matrix models
with cross LL interactions are also possible, although suppressed by the large $|eH|$
gap between LL.

\section{Ergodic Regime with Strong  Magnetism.}


For strong  QED magnetic fields with a running coupling
$g(eH)$ that is not so small, the semi-classical configurations such as instantons
(vacuum) or calorons (thermal) are not suppressed. Although 4-dimensional, these configurations may
considerably interact with the dimensionally reduced quark fields ${\bf \Psi}_2$ and
localize them in near zero modes. A recent discussion of these near zero modes for a single
instanton configuration can be found in~\cite{Basar:2011by}.
Let $N$ be the number of these near-zero modes in the transverse hyperplane to the
magnetic field. Since the chiralities in 4 dimensions  are commensurate with the
chiralities in 2 dimensions, the pertinent
chiral RMM for these near-zero modes is then standard

\begin{eqnarray}
&&{\bf Z}_{LLL}(m,T,\Lambda)=\nonumber\\
&&\int\,d{\bf A}\,e^{-(N/\Lambda^2){\rm Tr}\left({\bf A^\dagger A}\right)}
\,\,{\rm det}
\left|\begin{matrix}
m & i\omega_0+i{\bf A}\\
i\omega_0+i{\bf A}^\dagger& m\\
\end{matrix}\right|\nonumber\\
\label{18}
\end{eqnarray}
with ${\bf A}$ an $N\times N$ matrix  and $\Lambda\approx 1/\rho$ a typical QCD scale set by the instanton size $\rho$
for instance.
The gaussian weighting in (\ref{18}) induces a 4-Fermi interaction
$$\Lambda^2\left({\bf \Psi}_2^\dagger{\bf \Psi}_2\right)^2$$
for the dimensionally reduced quarks.
Note that the level spacing of the ${\bf \Psi}_2$ is

\be
\Delta\lambda\sim\frac{\Lambda}{N}
\label{18X}
\ee
instead of (\ref{12X}). Again, by the Banks-Casher formula we expect the chiral condensate in
a symmetric box to scale as

\be
\left<\Psi^\dagger_4\Psi_4\right>=\pi\rho(0)=\frac{\pi}{V_4} \,\frac{2\phi}{\Delta\lambda}
\sim\frac{|eH|}{\Lambda}\frac{N}{\sqrt{V_4}}
\label{18XX}
\ee
with ${\bf n}_4=N^2/V_4\approx 1/{\rm fm}^4$ as typically the instanton-anti-instanton density in the QCD vacuum.

The chiral condensate follows through similar reasoning at the saddle point for large $N$ and at finite temperature

\be
\left<\Psi^\dagger_4\Psi_4\right>=\frac{2N}{\beta L_z}\,\frac{|eH|}{\Lambda}\sqrt{1-\omega_0^2/\Lambda^2}
\label{14}
\ee
$N/\beta L_z$ is about the number of instantons and anti-instantons in
$\beta L_z$ regarded as a slice of $V_4=\beta L_xL_yL_z$. This
is roughly $\sqrt{\bf n}_4$ since the instanton density ${\bf n}_4$
varies slowly with temperature except for $T\approx T_c$ where a
re-arrangement into pairs is expected. Thus, the spacing
(\ref{18X}) between the eigenvalues is about $\Delta\lambda\approx 1/\sqrt{V_4}$
in the presence of a strong magnetic field instead of $\Delta\lambda\approx 1/V_4$
in the vacuum.

The chiral condensate is seen to grow linearly with the magnetic field.
Unlike the precedent matrix model, the transition temperature is
independent of the magnetic field

\be
T_c=\Lambda/\pi
\label{X14X}
\ee
Therefore, the slope of the chiral condensate for strong magnetism is

\be
\left<\Psi^\dagger_4\Psi_4\right>\approx \frac{2\sqrt{\bf n}_4}{\pi T_c}\,{|eH|}\,\sqrt{1-(T/{T_c})^2}
\label{14X}
\ee
In the RMM, the chiral condensate at zero temperature is fixed by similar arguments. As a result, the
relative change of the chiral condensate with magnetism is

\be
\Delta\Sigma=\frac{\left<\Psi^\dagger_4\Psi_4\right>}{\left<\Psi^\dagger_4\Psi_4\right>_0}\approx
\frac{{|eH|}}{\sqrt{\bf n}_4}\,\sqrt{1-(T/{T_c})^2}
\label{X14XX}
\ee
with ${\bf n}_4\approx 1/\,{\rm fm}^4$ and $T_c$ about the critical temperature for chiral symmetry restoration
without magnetism.

We note that the effects of the higher LL on the quark condensate can be readily assessed
in the generalized random matrix model (\ref{16X}) in the large $\phi$ (ultra-strong magnetism) and
large $N$ (strong magnetism) limit using the saddle point approximation.
Indeed, in the latter limit
and retaining the LLL $n=0, s=1$ along with the next to lowest LLL $n=0,s=-1$ and $n=1, s=1$ yield

\begin{eqnarray}
&&\left<\Psi^\dagger_4\Psi_4\right>=\frac{N}{\beta L_z}\,\frac{2|eH|}{\Lambda}\nonumber\\
&&\times\left(\sqrt{1-\omega_0^2/\Lambda^2}+2\sqrt{1-2|eH|/\Lambda^2-\omega_0^2/\Lambda^2}\right)\nonumber\\
\label{X14XXX}
\end{eqnarray}
The next to LLL contribution drops for strong fields or $|eH|>\Lambda^2/2$ as its corresponding
saddle point moves off the real axis. Typically $\Lambda\approx 0.5\,{\rm GeV}$ (see below) so that the decoupling takes
place for $|eH|\geq 1/3\,{\rm GeV}^2$ which is the onset of the ergodic regime. Thus our reduction to the LLL.

\section{Comparison with Lattice Simulations}

For intermediate values of $|eH|\leq 1\,{\rm GeV}^2$ our results for strong magnetism using random matrix theory
and zero temperature are overall  consistent with those reported by the recent lattice simulations in~\cite{{Bali:2013cf}}.
The deviations at low values of $|eH|<1/3\,{\rm GeV}^2$ occur in the diffusive regime and are determined by the pion
modes as we detailed earlier through the quark return probability. For ultra-strong magnetism our results appear to be
consistent with the lattice results reported in~\cite{DElia:2011zzb}.

At finite temperature, our results for strong magnetism using random matrix theory show a field independent critical
temperature (\ref{X14X}). The lattice results reported in~\cite{{Bali:2013cf}} show a critical temperature
weakly dependent on the magnetic field. The critical temperature appears to decrease by about 10\% across the
critical temperature. An anti-catalysis with a substantial decrease of the chiral condensate
was reported to take place close to the critical temperature.

In a recent investigation~\cite{Bruckmann:2013oba} the breaking of $Z_{N_C}$ symmetry at high temperature with
the appearance of a finite Polyakov holonomy was suggested as the mechanism for the appearance of the anti-catalysis
phenomenon in the lattice data.  The argument is that the finite temporal holonomy $A_4=\pi\varphi_4\,T$ with
$\varphi_4=(0,\pm 2/3)$ (mod 2) for $N_C=3$ or $\varphi_4=0,1$ (mod 2) for $N_C=2$ gets shifted by the magnetic field through sea
effects. The result is a decrease of the chiral condensate around the critical temperature.
In other words while the valence contribution is diamagnetic, the sea contribution through the Polyakov
holonomy appears to be paramagnetic.

Indeed, a small  shift in the trivial Polyakov holonomy by the applied magnetic field can cause the chiral
condensate to decrease substantially near the critical temperature. The effect of the Polyakov holonomy is to
alter the twisting of the temporal fermionic boundary condition and therefore an up or down shift in the Matsubara
frequencies. In the LLL the Polyakov holonomy is readily inserted through

\begin{eqnarray}
&&{\bf Z}_{LLL}(m,T,A_4)=\int\,d{\bf A}\,e^{-(N/\Lambda^2){\rm Tr}\left({\bf A^\dagger A}\right)}\nonumber\\
&&\times{\rm det}
\left|\begin{matrix}
m & i(\omega_0+A_4)+i{\bf A}\\
i(\omega_0+A_4)+i{\bf A}^\dagger& m\\
\end{matrix}\right|\nonumber\\
\label{T2}
\end{eqnarray}
instead of (\ref{18}). The immediate effect of the Polyakov holonomy is to change the onset of the
critical temperature to $T_c=(\Lambda/\pi)/|1+\varphi_4|$ (mod 2) in agreement
with the result in~\cite{Stephanov:1996he}.

The magnetic shift of the effective potential for the trivial Polyakov holonomy 
maybe estimated. At high temperature and for massless quarks,  the shift in the pressure 
due to magnetism is~\cite{Persson:1996zy} (see also section 6)

\be
\Delta \Omega_H\approx \frac {\chi}{2}|eH|^2{\rm ln}\left(\frac{\omega_0^2}{|eH|}\right)
\label{H1}
\ee
with $\chi=N_C/12\pi^2$. In the presence of a finite holonomy, (\ref{H1}) shifts through
$\omega_0\rightarrow \omega_0+A_4$ (mod 2) with additional contributions that are
not important for our arguments (see in general~\cite{Meisinger:1997jt}). 
%
As a result, the Polyakov potential around the trivial holonomy (mod 2)  reads

\be
\Delta\Omega_4\approx \frac 12m_E^2A_4^2+\frac {\chi}{\omega_0}\,A_4\,|eH|^2
\label{H3}
\ee
where we have added the leading contribution for zero magnetism.
Here $m_E^2$ is the electric screening mass. Without the electromagnetic shift, the
minimum of (\ref{H3}) is around the trivial holonomy $\varphi_4=A_4/\omega_0=0$
(mod 2). The magnetic contribution causes this minimum to shift

\be
\varphi_4(H)\approx 0-\chi\frac{|eH|^2}{(m_E\,\omega_0)^2}
\label{H4}
\ee
where the corrections are ${\cal O}(e^3)$. 
We note that around the critical temperature
in QCD with $N_C=3$, the electric mass $m_E\approx \pi T$ and $\chi=1/4\pi^2$. So
for $T_c\approx 0.160\,{\rm GeV}$ the Polyakov holonomy is shifted by about 10\%
for $|eH|\approx(\pi\,T_c)^2\approx (0.5\,{\rm GeV})^2$.

The shift (\ref{H4}) is rooted in the thermal cutoff in (\ref{H1}) through our
perturbative estimate. Let us assume that generically the trivial holonomy
shifts in the presence of magnetism with $\varphi_4(H)\approx {\cal O}(e^2)$
as also suggested by current lattica data~\cite{Bruckmann:2013oba}. While 
such a shift is small it can dramatically affect the chiral condensate near the
critical temperature. Indeed, inserting (\ref{H4}) in the random matrix model yields 
a chiral condensate

\begin{eqnarray}
\left<\Psi^\dagger_4\Psi_4\right>\rightarrow &&\frac{N}{\beta L_z}\,\frac{2|eH|}{\Lambda}\\
&&\times\frac12\sum_{s=\pm 1}
\sqrt{1-(T/T_c)^2(s+\varphi_4(H))^2}\nonumber
\label{shiftx}
\end{eqnarray}
The sum over $s=\pm$ signs in (\ref{shiftx}) is important to understand in our simplified random matrix theory.
The chiral condensate in the saddle point approximation receives contribution from {\bf all}
Matsubara frequencies $\omega_n=(2n+1)\pi T$ requiring in principle a matrix model with all
frequencies~\cite{book,Stephanov:1996he}. Such a matrix model can be written and is explicitly
periodic in $\varphi_4$ of period 2. However, near the critical points only the 2 lowest Matubara
frequencies are dominant, i.e.  $\omega_0=-\omega_{-1}$ (high temperature reduction). In the absence of holonomies
the random matrix model does not distinguish between these 2 frequencies
near the critical points, thus the
simplification to $\omega_0$. For a finite Polyakov holonomy, the 2 Matsubara
frequencies are distinguishable, thus the sum over the signs.

In the random matrix model the reduction of the chiral condensate takes place when

\be
\frac{T_c(H)}{T_c(0)}=\frac 1{|s+\varphi_4(H)|}\approx  1-s\varphi_4(H)
\label{Z14Z}
\ee
where the subscript refers to the critical point. The $\omega_{-1}$ contribution in (\ref{shiftx})
drops first for a critical temperature $T_c(H)/T_c(0)\approx 1-|\varphi_4(H)|$. With the earlier estimates
for $|eH|\approx (0.5\,{\rm GeV})^2$ the change in the critical temperature $T_c(H)$ with
the magnetic field is about 10\% lower than without,
in qualitative agreement with the lattice data~\cite{{Bali:2013cf}}. As a result (\ref{shiftx}) reduces at the critical point to

\be
\left<\Psi^\dagger_4\Psi_4\right>\rightarrow
\frac{N}{\beta L_z}\,\frac{|eH|}{\Lambda}\sqrt{|\varphi_4(H)|}\,\left(1-|\varphi_4(H)|\right)
\label{X14Z}
\ee
which is small and vanishes continuously with the increase of magnetism.
We note that (\ref{Z14Z}) implies a critical $H_c(T)$ through the perturbative
estimate

\be
eH_c(T)\approx \frac{(m_E\omega_0)_c}{\sqrt{\chi}}\,\left(1-\frac {T}{T_c}\right)^{1/2}
\label{14ZZ}
\ee
for the restoration of chiral symmetry at fixed magnetism and finite temperature for $N_c=3$. This is
in qualitative agreement with the reported lattice data. The case of $N_c=2$ will be adressed in a sequel.

As a final note, we would like to indicate that there is yet another way to test the nature
of magnetism on the lattice and that is by twisting  the fermionic boundary condition along $L_z$
with a Bohm-Aharonov flux $A_z$. Specifically

\be
\Psi_2(x_0, z+L_z)=-e^{i2\pi\varphi_z}\Psi_2(x_0,z)
\label{T1}
\ee
This amounts to
a bulk Abelian potential~\cite{Janik:1998jc,Nishigaki:2012rn,Stephanov:1996he}

\be
\left(i\gamma\cdot\nabla_E+A_z\gamma_{z}\right)\Psi_4=\lambda[A_z]\Psi_4
\label{XT1}
\ee
with a constant $A_z=2\pi\varphi_z/L_z$.  Recall that a constant $A_z$ is physical on a torus.
Note that for (\ref{XT1}) the quark virtualities $\lambda[A_z]$ are periodic in
$\varphi_z$ of period 1.
The twisted boundary condition (\ref{T1}) on the LLL in the random matrix model
is achieved by using

\begin{eqnarray}
{\rm det}
\left|\begin{matrix}
m & i(\omega_0+A_4)+A_z+i{\bf A}\\
i{\bf A}^\dagger+i(\omega_0+A_4)-A_z& m\\
\end{matrix}\right|\nonumber\\
\label{XT2}
\end{eqnarray}
in (\ref{T2}).
The chiral condensate follows through similar reasoning at the saddle point for large $N$ with

\be
\left<\Psi^\dagger_4\Psi_4\right>=\frac{N}{\beta L_z}\,\frac{2|eH|}{\Lambda}\sqrt{1-A_z^2/\Lambda^2-(\omega_0+A_4)^2/\Lambda^2}
\label{T3}
\ee
instead of (\ref{14}). At zero temperature, the change in the chiral condensate is

\be
\Delta\Sigma=\frac{\left<\Psi^\dagger_4\Psi_4\right>}{\left<\Psi^\dagger_4\Psi_4\right>_0}=
\frac{|eH|}{\sqrt{\bf n}_4}\,\sqrt{1-A_z^2/\Lambda^2}
\label{T4}
\ee


It would be interesting to carry numerical simulations of spectra and derive the chiral condensate from
the pertinent spectral fluctuations using directly the finite size random matrix construction with magnetism.
Sea effects can also be analyzed this way through relevant rescaling limits~\cite{Jurkiewicz:1996uy}, although the random matrix
model is totally dominated by the LLL with or without the Polyakov holonomy in the ergodic regime, as we
have discussed. The effects of light quark masses and $N_F>1$ will be addressed elsewhere.

\section{RMM Effective Action with Magnetism.}

The random matrix effective action with magnetism follows from (\ref{16X}) through standard fermionization
of the determinant and then bosonization by the use of the Hubbard-Stratonovitch transformation. The result
for the partition function (\ref{16X}) is

\be
Z_{LL}(m,T,\phi)=\int\,d{\bf B}\,e^{-\frac{N}{\Lambda^2}{\rm Tr}\,{\bf B}^\dagger{\bf B}}\
{\rm det}^N
\left|\begin{matrix}
{\bf B}+m & i{\bf L}\\
i{\bf L}^\dagger& {\bf B}^\dagger+m\\
\end{matrix}\right|
\label{Z1}
\ee
with $${\bf L}\equiv ((\omega_0+A_4)-i\epsilon(n,s))\,{\bf 1}_{ns}$$ a diagonal matrix labeling the Landau levels
shifted by the lowest Matsubara. The dominant contribution stems from the restriction to the LLL with
${\bf 1}_{01}$ and corresponds to the reduction ${\bf \Psi}_4$ to ${\bf \Psi}_2$ .
The addition of the spatial twist is straightforward. The corresponding effective action for the LLL
in 2-dimensions is

\begin{eqnarray}
\Omega_{LL}=\frac{N}{\Lambda^2}\,{\rm Tr}\,{\bf B}^\dagger{\bf B}
-N\,{\rm ln}\,{\rm det}\,\left|\begin{matrix}
{\bf B}+m & i{\bf L}\\
i{\bf L}^\dagger& {\bf B}^\dagger+m\\
\end{matrix}\right|
\label{Z2}
\end{eqnarray}
(\ref{Z2}) restricted to the LLL describes the dimensionally reduced effective action that describes
the quark spectrum near zero virtuality for strong fields in the ergodic regime. For instance

\begin{eqnarray}
\left<{\bf \Psi}_2^\dagger{\bf \Psi}_2\right>=\lim_{m\to0}\lim_{L_z\to\infty}
\frac{N}{\beta L_z}\left<{\rm tr}\,\left|\begin{matrix}
{\bf B}+m & i{\bf L}\\
i{\bf L}^\dagger& {\bf B}^\dagger+m\\
\end{matrix}\right|^{-1}\right>\nonumber\\
\label{Z3}
\end{eqnarray}
The averaging is carried using the measure in (\ref{Z1}).
The content of this dimensionally reduced effective action along with the role of a spatial twist and
a real quark chemical potential will be discussed elsewhere.

In what follows we will develop a more phenomenologically
inspired random matrix effective action that interpolates between the diffusive and ergodic regime
through a reduction of the NJL model to constant modes~\cite{book} (and references therein). The
results are inspired chiral RMM effective actions with magnetism. The purpose of this construction
is to show the potential relationship between the random matrix theory discussed above as suggested
by the mesoscopic analysis and the body of work on magnetism using NJL inspired models.

To analyze the chiral Euclidean effective action or pressure
we recall that the Gibbs energy  for finite T and $\mu$ of chiral RMM
is that of a two-level system for a particle and antiparticle for a single Matsubara frequency and
$N_F=1$~\cite{book,Jackson:1995nf}

\begin{eqnarray}
&&\Omega_M(T,\mu)=\frac 12 \Sigma^2 {\sigma^2}\nonumber\\&&
-\frac {{\bf n}_4}2\,{\rm ln}\left((\sigma-\mu)^2+(\pi T)^2)((\sigma+\mu)^2+(\pi T)^2)\right)\nonumber\\
\label{1}
\end{eqnarray}
with $\sigma$ playing the role of the constituent quark mass and $\Sigma^2$ the
strength of the chiral Gaussian noise. It ties to the chiral condensate in the vacuum through

\be
\left< \Psi_4^+\Psi_4\right>=\Sigma^2\sigma_*
\label{1X}
\ee
at the extremum.
Again, the density ${\bf n}_4={\bf N}/V_4$ refers in general to the number ${\bf N}$ of zero modes in the 4-volume which
is commensurate with the density of instantons plus antiinstantons in the vacuum.

AB fluxes act as Abelian U(1) vector potentials
and are readily  implemented in (\ref{1})  through the substitution

\be
(i\mu)^2\rightarrow (i\mu+A_4)^2+A_z^2
\label{2}
\ee
%
We recall that on a torus constant gauge fields are gauge invariant.
For $\mu=A_4=T=0$ (\ref{1}) simplifies

\be
\Omega_M(A_z)=\frac 12 \Sigma^2\sigma^2-\frac {{\bf n}_4}2\, {\rm ln}\left(\sigma^2+A_z^2\right)^2
\label{4}
\ee
with a chiral condensate at the minimum

\be
\left<{\bf \Psi}_4^\dagger{\bf \Psi}_4\right>\equiv\Sigma^2\sigma_*=
\Sigma\sqrt{2{\bf n}_4-(A_z\Sigma)^2}
\label{5}
\ee
that decreases with the twisting of the boundary condition as we noted earlier.


The virtuality levels of constituent quarks
in a fixed external magnetic field along the z-direction are fixed by (\ref{9}).
The vacuum energy follows by summing over the negative energy states through standard procedure.
Specifically, the negative of the energy per unit $V_4$ or pressure reads

\be
\Omega_H(0,0)=\frac{H^2}2+\frac{|eH|}{(2\pi)^2}\sum_{n=0}^\infty\int d k_z(2-{\delta}_{n0})\lambda^-_{n,1}(0,k_z)
\label{4XX}
\ee
where we have added the external magnetic field energy density contribution.
The sum is UV divergent and requires renormalization. Most importantly, we note that in the vacuum state
all the states  are paired in spin with the exception of the state $n=0$ and $s=1$. The disordered smearing
of this near-threshold state is responsible for the spontaneous breaking of chiral symmetry by strong magnetism
as we showed above.

The leading order in $1/N_c$ effects to the free constituent quark loop is the dominant contribution
whereby the constituent quark mass is generated through weaker residual and attractive interactions
say of the instanton type. For a fixed external magnetic field, the constituent quark loop contributes to
the magnetic permittivity $\mu (H)$. For a fixed cut-off $\Lambda_{UV}$ and a strong magnetic field or
$\Lambda_{UV}>\sqrt{|eH|}>(\sigma+m)$~\cite{Persson:1996zy}

\be
\Omega_H(0,0)\equiv \frac{H^2}{2\mu(H)}\approx \frac{H^2}{2}\left(1+\chi e^2\,{\rm ln}\left(\frac{\Lambda_{UV}^2}{eH}\right)\right)
\label{5XX}
\ee
for $N_F=1$. For several quark flavors $e^2\rightarrow \sum_q^{N_F}e_q^2$ with
$N_F=3$, $e_u=2e/3, e_d=-e/3, e_s=e/3$. Since the loop contribution stems from the UV sector, the
constituent quarks are for all purpose massless, making the logarithmic contribution unique.
The plus sign of the vacuum contribution reflects on the diamagnetism and hence the stability of the vacuum
state under the switching of a constant $H$. All the remaining contributions from the loop are analytic
in ${\cal O}((eH)^2)$ and go into the charge and $H$ renormalization. The combination $eH$ is renormalization
group invariant. We set the renormalization so that for $\Lambda_{UV}^2=|eH|$ (\ref{5XX}) is canonical~\cite{Persson:1996zy}.

\begin{figure}[t]
 \begin{center}
 \includegraphics[width=3cm]{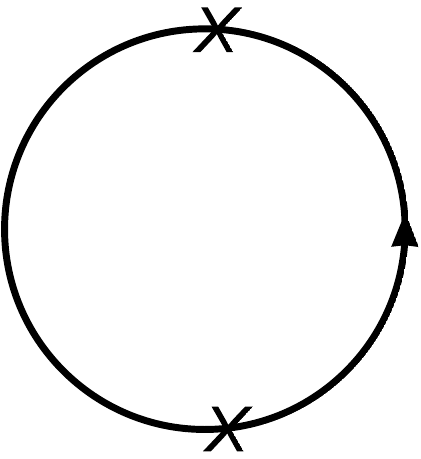}
 \caption{Magnetic contributions to the constitutive quark loop. See text.}
   \label{XLOOP}
 \end{center}
\end{figure}

The result (\ref{5XX}) is readily understood from the fermion loop diagram of Fig.~\ref{XLOOP} with the
fermionic lines corresponding to a constitutive fermion moving in the external magnetic field. The free
loop with 2 H-insertions is of the form $H^2/p^4$ with $p$ a typical loop momentum. Clearly the integral diverges both
in the ultraviolet and infrared.  The cutoff in the ultraviolet is $\Lambda_{UV}$ while in the infrared it is
$\sigma+m$ for free constitutive fermions and

\be
(\sigma+m)^2\rightarrow |eH|+(\sigma+m)^2
\ee
for constitutive fermions in a magnetic field.  For $\Lambda_{UV}>\sqrt{|eH|}>(\sigma+m)$ (\ref{5XX}) follows,
while for $\Lambda_{UV}>(\sigma+m)>\sqrt{|eH|}$ we have

\be
\Omega_H(0,0)\approx \frac{H^2}{2}\left(1+\chi e^2\,{\rm ln}\left(\frac{\Lambda_{UV}^2}{(\sigma+m)^2}\right)\right)
\label{5XXXbis}
\ee
For $\sqrt{|eH|}>(\sigma+m)$ we note that the higher H-insertions in Fig.~\ref{XLOOP} are increasingly
infrared sensitive making them all of order $|eH|^2$. They renormalize the classical field contribution
~\cite{Persson:1996zy}. In dense matter near the critical points, the UV cutoff in (\ref{5XXXbis}) is traded through

\be
\Lambda_{UV}^2\rightarrow \left|\omega_0+i\mu\right|^2
\label{5XXXX}
\ee
with $\omega_0=\pi T$, leading to

\be
\Omega_H(T,\mu)\approx \frac{H^2}{2}\left(1+\chi e^2\,{\rm ln}\left(\frac{\left|\omega_0+i\mu\right|^2}{|eH|+(\sigma+m)^2}\right)\right)
\label{5XXX}
\ee
in agreement with detailed limits derived in~\cite{Persson:1996zy}. (\ref{5XXX}) permeates all constitutive quark
calculations in the context of the magnetic catalysis of chiral symmetry breaking in QCD both in vacuum and matter
~\cite{Shovkovy:2012zn} (and references therein).


Our starting point for a RMM analysis of the phase diagram
at finite $H$ in matter with twisted boundary conditions is $\Omega=\Omega_M+\Omega_H$.
With only $H$ present, the minimum $\partial\Omega/\partial\sigma_*=0$ using (\ref{1}) and
(\ref{5XXX}) yields

\be
\sigma_*=\Sigma\sqrt{2{\bf n}_4+\chi H^2}
\label{8}
\ee
which is to be compared with (\ref{5}).  This yields a change in the chiral condensate

\be
\Delta\Sigma=\frac{\left<\Psi^\dagger_4\Psi_4\right>}{\left<\Psi^\dagger_4\Psi_4\right>_0}=
\sqrt{1+\frac{\chi\,|eH|^2}{2{\bf n}_4}}-1
\label{9X}
\ee
which is quadratic for low $|eH|$ and linear for intermediately large $|eH|$. We recall that for ultra-strong
magnetism at zero temperature the ratio asymptotes $|eH|^{3/2}$ as both suggested by RMM and perturbative QCD.

The asymptotic of (\ref{9X})

\be
\Delta\Sigma\approx
\sqrt{\frac{(\chi/2)}{{\bf n}_4}}\,\,|eH|
\label{10X}
\ee
is smaller than (\ref{X14XX}) by the factor
\be
\sqrt{\chi/2}\approx \sqrt{N_C/24}/\pi\approx 1/10
\label{FACTOR}
\ee
The random matrix model  is
not restricted by the 1-loop result as it involves the delocalization of the dimensionally
reduced chiral quarks through quasi-2-dimensional zero modes and is more in line
with the currently reported lattice results.  The results of the random matrix analysis are
reproduced by setting $\chi\rightarrow 2$ in $\Omega_H$. This effective action
analysis without boundary twists whether temporal (Polyakov) or spatial (BA),
emphasizes the underlying diamagnetic character of the random matrix analysis above
when restricted to the LLL.


\section{Conclusions and Prospects}

We have presented a chiral RMM for the analysis of the recent lattice QCD
simulations in the presence of a strong QED magnetic field. The results for intermediate values of
$|eH|>1/\rho^2\approx 0.3\,{\rm GeV}^2$ are readily understood through the infrared branch of the LLL which causes the
quarks to dimensionally reduce from 4 to 2 dimensions. Quarks trapped in 4 dimensional LLL
are more prone to break spontaneously chiral symmetry through random disorder by gluonic
configurations whether perturbative or non-perturbative. This is the essence of the magnetic catalysis
discussed in~\cite{Shovkovy:2012zn}
(and references therein). Note that our reduction of the spectrum is in quark virtualities
not quark energies and therefore from $4+1$  to $2+1$ rather than $3+1$ to $1+1$.

In 2 dimensions the interactions due to
the semiclassical vacuum configurations, i.e. instantons or calorons cause them to accumulate
at zero virtuality. The accumulation is characterized by a level spacing of order $1/\sqrt{V_4}$
which is intermediate between $1/V_4$ in the vacuum and $1/{}^4\sqrt{V_4}$ in a free box.
The change in the chiral condensate is shown to increase linearly with $|eH|$ (catalysis) with a slope
$1/\sqrt{{\bf n}_4}\approx 1\,{\rm fm}^2$ with ${\bf n}_4$ the instanton vacuum density. These
results are consistent with the current lattice data at zero temperature. For very weak $|eH|$ the gapped charged pions
dominate the contribution in the diffusive regime in line with the leading result from chiral
perturbation theory. At ultra-strong magnetic fields $|eH|\geq 10/\rho^2\approx 3\,{\rm GeV}^2$
the RMM suggests an increase in the chiral
condensate as $|eH|^{3/2}$ in agreement with
perturbative QCD estimates for ultra-strong fields using
a Bethe-Salpeter analysis~\cite{Shushpanov:1997sf}.

At finite but high temperature the emergence of a Polyakov holonomy does not affect
the decoupling of the LLL. We have shown that a small  shift in the trivial part
of this holonomy  allows for a rapid decrease (anti-catalysis) of the chiral condensate near the critical
temperature as suggested recently by lattice simulations~\cite{Bruckmann:2013oba}.
This small change in the Polyakov holonomy is quadratic in the magnetic field.
It reflects on the back-reaction of the quarks on the effective potential of the temporal
holonomies near the trivial solution.

We have explicitly shown how the random matrix effective action emerges from our analysis
in the dimensionally reduced limit. Also, we have provided a specific construction of a number
of random matrix inspired models to allow for a simple comparison with current constituent
quark models with magnetism such as the NJL models. A more detailed analysis of the phase
diagrams emerging from these effective actions will be detailed elsewhere.

In random matrix theory the spectral distributions in the microscopic
limit capture more information on the subtleties of chiral symmetry breaking
such as the role of quark representations, quark masses and magnetism and
twists.

Of particular interest is $N_C=2$ with an external magnetic field which
upsets time-reversal symmetry, a big deal for this representation. Indeed,
for $N_C=2$  the chiral disorder involves
not only the pions (diffusons) but also the baryons (cooperons) both of which are degenerate
following the extra Pauli-Gursey symmetry. The cooperon is sensitive to the breaking of
time-reversal symmetry because of its charge~\cite{Janik:1998jc}.

Finally, it would be interesting to revisit the role of the number of flavors, quark masses
and the spatial and temporal twists on the results we have derived for a more critical
comparison with current lattice data. Also the effect of the chemical potential on the LLL
analysis should shed more light on the role of magnetism on the QCD phase diagram.
Some of these issues will be addressed next.


\section{ Acknowledgements.}

IZ would like to thank Dima Kharzeev and Edward Shuryak for discussions. MAN appreciates early discussion  with Waldemar Wieczorek on the role of the magnetic field in the diffusive regime.  
IZ is supported in part  by the U.S. Department of Energy under Contract No.
DE-FG-88ER40388.
MAN is  supported in part by the Grant DEC-2011/02/A/ST1/00119 of the National Centre of Science.



\begin{thebibliography}{99}
\bibitem{DIAKONOV}
D. Diakonov, Prog. Part. Nucl. Phys. {\bf 51}, 173 (2003).
\bibitem{Schafer:1996wv}
  T.~Schaefer and E.~V.~Shuryak,
  Rev.\ Mod.\ Phys.\  {\bf 70}, 323 (1998)
  [hep-ph/9610451].


\bibitem{book}
M.~Nowak, M.~Rho and I.~Zahed,
Chiral Nuclear Dynamics, World Scienfific, Singapore (1995). 


\bibitem{Banks:1979yr}
  T.~Banks and  A.~Casher,
  Nucl.\ Phys.\ B {\bf 169}, 103 (1980).

\bibitem{Janik:1998ki}
  R.~A.~Janik, M.~A.~Nowak, G.~Papp and I.~Zahed,
  Phys.\ Rev.\ Lett.\  {\bf 81}, 264 (1998)
  [hep-ph/9803289].
\bibitem{OSVERBAAR}
J.C. Osborn and J.J.M. Verbaarschot, Phys. Rev. Lett. {\bf 81}, 268 (1998). 
\bibitem{TAKEH}
K. Takahashi and I. Iida, Nucl. Phys. {\bf B573}, 685 (2000). 
\bibitem{BERBE}
M. E. Berbenni, T. Guhr, J.Z. Ma, S. Meyer et al.,  Nucl. Phys. Proc. Supp. {\bf 83}, 914 (2000). 
\bibitem{Shovkovy:2012zn}
  I.~A.~Shovkovy,
  arXiv:1207.5081 [hep-ph].

\bibitem{McLerran:2003yx}
  L.~McLerran,
  hep-ph/0311028.


\bibitem{Bali:2013cf}
G.~S.~Bali, F.~Bruckmann, G.~Endrodi, Z.~Fodor, S.~D.~Katz {\it et al.}, JHEP \textbf{1202}, 044 (2012) [arXiv:1111.4956],
 Phys. Rev. \textbf{D86}, 071502 (2012) [arXiv:1206.4205]; G.~S.~Bali, F.~Bruckmann, M.~Constantinou, M.~Costa, G.~Endrodi {\it et al.},
 Phys. Rev. \textbf{D86}, 094512 (2012) [arXiv:1209.6015];
 G.~S.~Bali, F.~Bruckmann, M.~Constantinou, M.~Costa, G.~Endrodi, Z.~Fodor, S.~D.~Katz and S.~Krieg {\it et al.},
  arXiv:1301.5826 [hep-lat].

\bibitem{Bruckmann:2013oba}
  F.~Bruckmann, G.~Endrodi and T.~G.~Kovacs,
  arXiv:1303.3972 [hep-lat].

\bibitem{Shushpanov:1997sf}
  I.~A.~Shushpanov and A.~V.~Smilga,
  Phys.\ Lett.\ B {\bf 402}, 351 (1997)
  [hep-ph/9703201].



\bibitem{Janik:1998cg}
  R.~A.~Janik, M.~A.~Nowak, G.~Papp and  I.~Zahed,
  Acta Phys.\ Polon.\ B {\bf 29}, 3215 (1998)
  [hep-ph/9807467].

\bibitem{Gasser:1987ah}
  J.~Gasser and H.~Leutwyler,
  Phys.\ Lett.\ B {\bf 188}, 477 (1987);
  J.~Gasser and H.~Leutwyler,
  Phys.\ Lett.\ B {\bf 184}, 83 (1987).

\bibitem{Jackson:1995nf} 
  A.~D.~Jackson and J.~J.~M.~Verbaarschot,
  Phys.\ Rev.\ D {\bf 53}, 7223 (1996)
  [hep-ph/9509324].

\bibitem{Basar:2011by}
  G.~Basar, G.~V.~Dunne and  D.~E.~Kharzeev,
  Phys.\ Rev.\ D {\bf 85}, 045026 (2012)
  [arXiv:1112.0532 [hep-th]].

\bibitem{DElia:2011zzb}
  M.~D'Elia, S.~Mukherjee and F.~Sanfilippo,
  AIP Conf.\ Proc.\  {\bf 1343}, 525 (2011).




\bibitem{Jurkiewicz:1996uy}
  J.~Jurkiewicz, M.~A.~Nowak and I.~Zahed,
  Nucl.\ Phys.\ B {\bf 478}, 605 (1996)
  [Erratum-ibid.\ B {\bf 513}, 759 (1998)]
  [Nucl.\ Phys.\ B {\bf 513}, 759 (1998)]
  [hep-ph/9603308].

\bibitem{Janik:1998jc}
  R. A. Janik, M. A. Nowak, G. Papp, and I. Zahed, Phys. Lett. {\bf B 440}, 123 (1998); arXiv: hep-ph/9807499; Nucl. Phys.
Proc. Suppl. 83, 977 (2000).
  
\bibitem{Nishigaki:2012rn} 
  S.~M.~Nishigaki,
  Phys.\ Rev.\ D {\bf 86}, 114505 (2012)
  [arXiv:1208.3452 [hep-lat]].


\bibitem{Persson:1996zy}
  D.~Persson,
  Annals Phys.\  {\bf 252}, 33 (1996)
  [hep-ph/9601259].


\bibitem{Meisinger:1997jt} 
  P.~N.~Meisinger and M.~C.~Ogilvie,
  Phys.\ Lett.\ B {\bf 407}, 297 (1997)
  [hep-lat/9703009].



\bibitem{Stephanov:1996he}
  M.~A.~Stephanov,
  Phys.\ Lett.\ B {\bf 375}, 249 (1996)
  [hep-lat/9601001].




\end{thebibliography}
\end{document}